\documentclass[10pt,twocolumn,preprintnumbers,amsmath,amssymb,nofootinbib,superscriptaddress]{revtex4-1}

\usepackage{amsmath}  
\usepackage{amssymb}  
\usepackage{theorem}

\usepackage{graphicx}
\usepackage{dcolumn}
\usepackage{amssymb,amsmath,bm}
\usepackage{color}
\usepackage[colorlinks,linkcolor=red,citecolor=blue,urlcolor=blue ]{hyperref}
\usepackage{multirow}
\usepackage{enumitem}
\usepackage{mathptmx} 
\usepackage{tensor}
\usepackage{amssymb} 
\usepackage{lettrine}
\usepackage[normalem]{ulem}
\usepackage{empheq}
\usepackage{comment}

\def \be {\begin{equation}}
\def \ee {\end{equation}}

\newcommand{\beq}{\begin{equation}}
\newcommand{\eeq}{\end{equation}}
\newcommand{\bea}{\begin{eqnarray}}
\newcommand{\eea}{\end{eqnarray}}


\usepackage[stable]{footmisc}

\newcommand\ees{\end{eqnarray}}
\newcommand\bees{\begin{eqnarray}}

\begin{document}
  \title{Chiral Gravitational Waves in Palatini Chern-Simons}
  
\author{Felipe Sulantay}
 \email{fmsulantay@uc.cl}
  \affiliation{Facultad de Fisica, Pontificia Universidad Cat\'olica de Chile, Av. Vicuña Mackenna 4860, Santiago, Chile}
   \author{Macarena Lagos}
 \email{m.lagos@columbia.edu}
  \affiliation{Department of Physics and Astronomy, Columbia University, New York, NY 10027, USA}
     \author{M\'aximo Ba\~nados}
  \email{maxbanados@fis.uc.cl}
  \affiliation{Facultad de Fisica, Pontificia Universidad Cat\'olica de Chile, Av. Vicuña Mackenna 4860, Santiago, Chile}

\begin{abstract}
We study the parity-breaking higher-curvature gravity theory of Chern-Simons (CS), using the Palatini formulation in which the metric and connection are taken to be independent fields. We first show that Palatini CS gravity leads to first-order derivative equations of motion and thus avoid the typical instabilities of CS gravity in the metric formalism. As an initial application, we analyze the cosmological propagation of gravitational waves (GWs) in Palatini CS gravity. We show that, due to parity breaking, the polarizations of GWs suffer two effects during propagation: amplitude birefringence (which changes the polarization ellipticity) and velocity birefringence (which rotates the polarization plane). While amplitude birefringence is known to be present in CS gravity in the metric formalism, velocity birefringence is not present in metric CS gravity for high frequency waves, but now appears in Palatini CS due to the fact that left-handed and right-handed GW polarizations have a different dispersion relation. In the approximation of small deviations from General Relativity (GR), we do find however that velocity birefringence appears at least  quadratically in the CS coupling parameter $\alpha$, while amplitude birefringence appears linearly in $\alpha$. This means that amplitude birefringence will be the most relevant effect in Palatini CS and hence this model will behave similarly to metric CS. We confirm this by applying current constraints on amplitude and velocity birefringence to Palatini CS, and showing that those from amplitude birefringence give the tightest bounds.
\end{abstract}

\keywords{Parity violation, gravitational waves}

\maketitle

\section{Introduction}\label{sec:intro}
The detection of gravitational waves (GWs) from binary compact objects mergers by the LIGO/Virgo collaboration \cite{LIGOScientific:2021djp} has opened the possibility of performing direct tests of gravity in the weak and strong-field regimes. Deviations from General Relativity (GR) in the generation and propagation of GWs have been analyzed, and used to constrain various modified gravity models and to probe fundamental physics \cite{Baker:2017hug, LIGOScientific:2018dkp, LIGOScientific:2020tif, Ghosh:2021mrv, LIGOScientific:2021sio}. 

One well-studied model is Chern-Simons (CS) gravity \cite{Deser:1981wh, Campbell:1990fu, Lue:1998mq, Jackiw:2003pm}, which is a metric theory that contains parity-breaking quadratic curvature terms coupled to a scalar field. As a consequence, in this metric formulation, the CS theory contains third-order derivatives of the metric in the equations of motion, and is likely to be ill-posed \cite{Delsate:2014hba}, so it must be considered as an effective field theory. A perturbative reduction scheme \cite{Okounkova:2017yby} has been developed in order to recast this theory in terms of first-order derivative equations and shown to have a well-posed boundary problem (with a unique stable solution continuous on the initial data), and hence allow for non-linear numerical GW simulations \cite{Okounkova:2019dfo, Okounkova:2022grv}.

An alternative formulation to Chern-Simons was proposed in \cite{BottaCantcheff:2008pii} in terms of a tetrad and spin connection, with an internal Minkowski metric. This model only leads to first-order derivative equations of motion, and hence does not suffer from the same issues as the original CS metric formulation. Note that the metric and tetrad CS formulations are generally inequivalent since the tetrad formulation typically leads to a non-vanishing torsion in the curvature. Up to date, most physical predictions have been studied for CS in the metric formalism, so the viability of the tetradic CS model still remains to be studied. 

In this paper, we present a third alternative formulation of CS, based on the so-called Palatini formulation \cite{1982GReGr..14..243F}, in which the spacetime metric and the connection are assumed to be independent fields. Palatini formulations have been identified in the past \cite{Olmo:2011uz} as a way of avoiding instabilities induced by high curvature interactions, since they avoid higher derivative equations of motion. In this paper, we calculate the full nonlinear equations and show they contain only first-order derivatives. Thus, Palatini CS avoids the typical instabilities present in the metric CS model. Contrary to the tetradic CS formulation, here we assume the connection to be always a symmetric tensor, which means that the curvature will always be torsion free. Note that an extension to Chern-Simons gravity \cite{Boudet:2022wmb, Boudet:2022nub, Bombacigno:2022naf} was recently analyzed using the Palatini formulation as well, and this model was found to contain metric torsion.

Beside the instability issues there is a purely theoretical motivation to study Palatini's formulation of gravity: from a differential geometry point of view the connection and metric are different concepts. The metric is associated to lengths of  curves and norms of vectors, while the connection is associated to parallel transport. It is then reasonable to treat them as independent entities from a dynamical point of view. This exercise, when applied to the Einstein-Hilbert action, gives the usual Levi-Civita symbol, but this is not the case for other interactions, like Chern-Simons theory considered in this paper.

In addition, we analyze some of the physical predictions of Palatini CS, focusing on the propagation of GWs on an expanding Universe. In particular, we analyze linear tensor cosmological perturbations and we first find that the connection does not carry independent tensor degrees of freedom, and thus we can obtain a closed system of equations for the two GW polarizations, just like in the metric CS model. Nevertheless, the details of these equations differ from those in the metric CS model, which means that they predict a different phenomenology.

More concretely, in both Palatini and metric CS the two GW polarizations propagate differently due to parity breaking.  However, this parity breaking is exhibited only as an amplitude birefringence effect (which changes the relative amplitude of the two polarizations) in the metric CS model \cite{Lue:1998mq, Alexander:2007kv, Yunes:2010yf, Yagi:2017zhb, Okounkova:2021xjv}. Instead, in Palatini CS we find the presence of velocity birefringence (which changes the relative phase of the two polarizations due to modified dispersion relations) in addition to amplitude birefringence---see \cite{Zhao:2019xmm} for a discussion on velocity and amplitude birefringence of GWs in different gravity models. Furthermore, we find that the specific way in which amplitude birefringence happens in the Palatini and metric CS models is different in general.

In this paper we also discuss the prospects for constraining Palatini CS in the context of late-time cosmology. An initial constraint on amplitude birefringence has been obtained from the binary black hole mergers detected by LIGO/Virgo \cite{Okounkova:2021xjv}, and future binary neutron star mergers with electromagnetic counterparts could test this effect even further \cite{Yunes:2010yf}. We find that when the deviations from GR are small (i.e.\ the Chern-Simons interaction is small) then these constraints apply to both metric and Palatini CS since they both make the same prediction on amplitude birefringence at leading order. 

In addition, changes in the phase of GW polarizations have been previously tested while preserving parity symmetry. Constraints on the propagation speed of GWs \cite{LIGOScientific:2017zic} as well as on a modified dispersion relation have been obtained with LIGO/Virgo events \cite{LIGOScientific:2021sio}. Constraints on a specific parity-breaking modified dispersion relation have also been obtained in \cite{Wang:2020cub, Wang:2021gqm, Zhao:2022pun}. We find that when the deviations from GR are small, Palatini CS predicts a parity-preserving change in the propagation speed of GWs, while higher-order terms break parity and modify the GR dispersion relation. We thus discuss how the previous results from \cite{LIGOScientific:2017zic} and \cite{Zhao:2022pun} can be applied to Palatini CS, and generally lead to weaker bounds compared to those from amplitude birefringence.

This paper is structured as follows. In Section \ref{sec:CS} we review the Chern-Simons theory, and present the Palatini formalism while comparing to the metric formalism. In Section \ref{sec:cosmo} we analyze the cosmological predictions of Palatini Chern-Simons, focusing on the propagation of GWs over a homogeneous and isotropic expanding universe. We present the relevant GW equations of motion and compare to the metric formalism. Then, in Section \ref{sec:GWs} we reduce the metric and connection equations to a simple set of equations for the two physical polarizations of GWs, show that both amplitude and velocity birefringence are present, and discuss the observational prospects for constraining Palatini CS. Finally, in Section \ref{sec:discussion} we summarize our results and discuss future implications.

In this paper, we will set the speed of light to unity $c=1$ and $\hbar=1$. We will also use the $(-,+,+,+)$ metric signature, and denote with parenthesis the symmetrization of indices as: $A_{(\mu} B_{\nu)}= (A_{\mu}B_{\nu} + A_{\nu}B_{\mu})/2$.

\section{Chern Simons}\label{sec:CS}
General Relativity is an effective field theory (EFT) that breaks down at high energies \cite{Donoghue:1994dn, Burgess:2003jk}, which is reflected in the fact that it is a non-renormalizable theory whose high-energy quantum states cannot be analyzed. 
However, classical extensions to GR that include higher-order curvature terms have been proposed as a way to improve the validity of GR to higher energies, such as the case of Chern-Simons and $f(R)$ theories \cite{Sotiriou:2008rp, DeFelice:2010aj}. Such high curvature terms introduce their own problems since higher derivative terms lead to the presence of instabilities \cite{Stelle:1977ry, Nunez:2004ts, Chiba:2005nz}. However, they can also be considered as EFTs such that all the unstable degrees of freedom are excited at energies higher than the cutoff energy.

In this section, we start by reviewing the CS theory. Chern-Simons was first proposed as a modified gravity in \cite{Jackiw:2003pm} (see also a comprehensive review in \cite{Alexander:2009tp}). For a spacetime metric $g_{ab}$ and a pseudo scalar field $\vartheta$, the action is given by
\begin{align}
    S & = \kappa \int d^4 x \; \sqrt{-g} R  +  \frac {\alpha}{4} \int d^4 x \;\vartheta (\phantom{}^* R^{\alpha \: \mu \nu}_{\;\beta} R^{\beta}_{\;\alpha \mu \nu})  \;  \nonumber\\
    & -\frac{\beta}{2}\int \sqrt{-g} \; [g^{\alpha \beta}\;(\nabla_{\alpha}\vartheta)(\nabla_{\beta}\vartheta)+2V(\vartheta)]\; d^4x + \; S_{mat}, \label{CSaction}
\end{align}
where $\phantom{}^{*}R^{\alpha \: \mu \nu}_{\;\beta}$ is the Dual Riemann tensor defined as:
\begin{equation}
    \phantom{}^* R^{\alpha \: \mu \nu}_{\;\beta}=\frac{1}{2}\varepsilon^{\mu \nu \rho \sigma} R^{\alpha}_{\;\beta \rho \sigma},
\end{equation}
with $\varepsilon^{\mu \nu \rho \sigma}$  the fully anti-symmetric Levi-Civita symbol.
In Eq.\ (\ref{CSaction}), the first term corresponds to the Einstein-Hilbert action of GR. The second term contains the Chern-Simons interaction---also known as the Pontryagin Density---and depends on the Levi-Civita symbol $\varepsilon^{\mu \nu \rho \sigma}$  which will induce parity breaking in the solutions. Note that this CS term contains higher-curvature interactions but it can affect the behavior of gravity in the weak-field regime of cosmology as well, depending on the scales involved in the problem. Note also that, in Palatini form, this term is independent from the metric tensor, yet is has the right transformation properties thanks to the Levi-Civita density. The third term in Eq.\ (\ref{CSaction}) is the kinetic action for the scalar field $\vartheta$ that includes an arbitrary potential term $V$, and lastly $S_{mat}$ includes additional general matter components, assumed to be minimally coupled to the metric, as in GR. The parameters $\alpha$ and $\beta$ are arbitrary real dimensional coupling constants characterizing the scalar field, and $\kappa=1/(16\pi G)$. 

\subsection{CS: metric formulation}\label{subsec_2ndorder}
In the initial work \cite{Jackiw:2003pm}, CS was studied in the metric formalism, where there are two independent fields: the metric $g_{\mu\nu}$ and the scalar field $\vartheta$.
In this case, the metric equations of motion from Eq.\ (\ref{CSaction}) are
\begin{equation}
    \kappa G_{\mu \nu}  = - \alpha C_{\mu \nu} +\frac{1}{2}T_{\mu \nu}, \label{CSgeq}
\end{equation}
where $G_{\mu \nu}=R_{\mu \nu}-g_{\mu \nu}R/2$ is the Einstein tensor which, in this formalism, is a function of the metric only. The C-tensor in the RHS of Eq.\ (\ref{CSgeq}) is given by:
\begin{equation}
    C^{\mu \nu}=(\nabla_{\rho}\vartheta)\epsilon^{\rho \sigma \lambda(\mu}\nabla_{\lambda}R^{\nu)}_{\;\sigma} +(\nabla_{\rho}\nabla_{\sigma}\vartheta) ^{*}R^{\sigma(\mu \nu)\rho}. \label{Ctensor}
\end{equation}
Also, $T_{\mu \nu}$ is the total stress-energy tensor, including $\vartheta$ and matter, defined as
\begin{align}
    T_{\mu \nu}&= T^{\vartheta}_{\mu \nu} + T^{mat}_{\mu \nu};  \label{st}\\
    T^{mat}_{\mu \nu} &\equiv  -\frac{2}{\sqrt{-g}} \frac{\delta \mathcal{L}_{mat}}{\delta g^{\mu \nu}}, \quad T^{\vartheta}_{\mu \nu} \equiv-\frac{2}{\sqrt{-g}}  \frac{\delta \mathcal{L}_{\vartheta}}{\delta g^{\mu \nu}},\nonumber
\end{align}
with
\begin{equation}
    T^{\vartheta}_{\mu \nu}=\beta \left(\nabla_{\mu}\vartheta\nabla_{\nu}\vartheta-\frac{1}{2}\nabla_{\lambda}\vartheta \nabla^{\lambda}\vartheta g_{\mu \nu}-V(\vartheta) g_{\mu \nu}\right),
\end{equation}
while $T_{\mu \nu}^{mat}$ will depend on the specific physical scenario of interest. We will assume that additional matter components do not interact with the scalar field $\vartheta$ and conserve independently:
\begin{equation}
\nabla^{\mu}T_{\mu \nu}^{mat}    =0. \label{Eq_mat_cons}
\end{equation}
We also obtain the following scalar field equation of motion:
\begin{equation}
  \beta \Box \vartheta -\beta \frac{dV(\vartheta)}{d\vartheta}=-  \frac{\alpha}{4}\:^{*}R R .\label{eq:theta2nd}
\end{equation}

As we can see from the C-tensor in Eq.\ (\ref{Ctensor}), the metric equation contains first-order derivatives of the Riemann tensor, and hence third-order derivatives of the metric. This means that Ostrogradski's ghost instabilities \cite{Ostrogradsky:1850fid, Aoki:2020gfv} may generically appear. The instability can be avoided only if one considers CS to be an EFT, such that the ghost only becomes relevant for energies higher than the cut-off scale of the EFT. If we restrict ourselves to physics below this cut-off, then the CS terms will always describe small deviations from GR. For this reason, similarly to previous works, in this paper we will make the small coupling $\alpha$ approximation when analyzing GWs in Section \ref{sec:GWs}. 

Furthermore, this same problem limits the numerical analysis of the full nonlinear equations in the metric CS model, and only a perturbative formulation has been proposed as a way of avoiding the instabilities \cite{Okounkova:2017yby, Okounkova:2019dfo}.  

\subsection{CS: Palatini formulation}\label{subsec_1CS_Eqns}
In a Palatini formulation, the relationship between the metric and the connection is determined dynamically by the equations of motion. In General Relativity (only an Einstein-Hilbert interaction) the Palatini and metric formulations give rise to the same metric-connection relation, namely, the Levi-Civita connection. When other interactions are added, for example, the Chern-Simons term, both variations lead to different results and different physics. In this section, we calculate the equations of motion of Chern-Simons gravity in the Palatini formulation, and show they are distinct from the metric CS model. Most importantly, in the Palatini variation no higher derivatives show up.

We will start by assuming the connection $\Gamma^\alpha
_{\mu \nu}=\Gamma^\alpha_{\nu \mu}$ to be symmetric (i.e.\ there is no curvature torsion), but this assumption could be relaxed in the future. In this theory, in Eq.\ (\ref{CSaction}) the Riemann tensor must be expressed directly as a function of the connection:
\begin{equation}
    R^{\alpha}_{\; \beta \mu \nu} = \partial_\mu\Gamma^\alpha_{\nu \beta}-\partial_\nu\Gamma^\alpha_{\mu \beta}+\Gamma^\alpha_{\mu \lambda}\Gamma^\lambda_{\nu \beta}-\Gamma^\alpha_{\nu \lambda}\Gamma^\lambda_{\mu \beta},
\end{equation}
and similarly for the covariant derivatives. The metric only enters in Eq.\ (\ref{CSaction}) by raising and lowering indices as well as through the determinant $g$.

In this formalism, there are three independent fields: the metric $g_{\mu\nu}$, the connection $\Gamma^\alpha_{\mu \nu}$, and the scalar $\vartheta$. For the metric tensor, we obtain
\begin{equation}
    G_{\mu \nu}=\frac{1}{2\kappa}T_{\mu \nu}, \label{Eqg_CS1}
\end{equation}
where now $G_{\mu \nu}$ is a function of the metric and the connection, while $T_{\mu \nu}$ is the same as \eqref{st}. We can see that in this formalism the C-Tensor does not appear in the equation for the metric because the Chern-Simons term does not longer depend on the metric tensor. Indeed, Eq.\ (\ref{Eqg_CS1}) has the same form as the equation for General Relativity.

Then, we obtain the equation for the scalar field 
\begin{equation}
    \beta\Box_{g}\vartheta -\beta\frac{d V(\vartheta)}{d \vartheta}  =- \frac{\alpha}{4}\;R^{*}R, \label{eq:theta1st}
\end{equation}
where $\Box_{g}$ denotes the metric-compatible D'Alambertian operator
\begin{equation}
    \Box_{g}\equiv {1\over \sqrt{-g}} \partial_{\alpha}(\sqrt{-g}g^{\alpha \beta}\partial_{\beta}).
\end{equation}
Note that Eq.\ (\ref{eq:theta1st}) is the same equation as in the metric CS formalism (c.f.\ Eq.\ (\ref{eq:theta2nd})). 

Finally, the equation for the connection is 
\begin{align}
   &  \nabla_{\alpha}(\sqrt{|g|}g^{\beta \sigma}) - \frac{1}{2}\nabla_{\mu}(\sqrt{|g|}g^{\beta \mu})\delta^{\sigma}_{\alpha} - \frac{1}{2}\nabla_{\mu}(\sqrt{|g|}g^{\sigma \mu})\delta^{\beta}_{\alpha}\nonumber\\
   &  -\frac{1}{2}\alpha\; \varepsilon^{\mu \nu \rho \sigma} R^{\beta}_{\;\alpha \mu \nu}\nabla_{\rho}\vartheta -\frac{1}{2}\alpha\;  \varepsilon^{\mu \nu \rho \beta} R^{\sigma}_{\;\alpha \mu \nu}\nabla_{\rho}\vartheta =0	\label{conection},
\end{align}
where we have assumed, as usual for bosonic matter, that the additional matter components that may be present in the theory do not depend on the connection.
We see that the Chern-Simons interaction now affects the connection equation (\ref{conection}). In particular, we find that Eq.\ (\ref{conection}) depends on the scalar field $\theta$ whenever $\alpha\not=0$. As a consequence, the relationship between the metric and connection is expected to depend on $\theta$, contrary to the metric CS formulation with the Levi-Civita connection. However, if the scalar field vanished, $\theta=0$, then Eq.\ (\ref{conection}) would not depend on the CS interaction term, and would lead to:
\begin{equation}
    \nabla_{\mu}(\sqrt{|g|}g^{\mu \sigma})=0,
\end{equation}
which gives the Levi-Civita connection.

In addition, we find that now for the metric and the connection, all the equations of motion contain only first-order derivatives, and hence this formulation has the minimal required properties to ensure stability of the solutions. Whether this theory is free of other kind of instabilities and behaves generally well must be analyzed in detail in the future. For now, we focus on one concrete physical setup that is discussed in the next section.


\section{Cosmology}\label{sec:cosmo}
In this section we analyze the cosmological predictions of Palatini Chern-Simons. We will also review results on the metric CS model to compare and highlight their similarities and differences.

We first discuss the evolution of the cosmological background, and then move onto the behavior of linear cosmological perturbations. For simplicity, we analyze tensor perturbations only, and leave the analysis of scalar and vector perturbations for future work.

\subsection{Background}
Motivated by current cosmological observations, let us assume an isotropic and homogeneous spatially-flat expanding Universe that describes the average behavior on large scales. The spacetime element is then described by the Friedmann-Robertson-Walker (FRW) metric as:
\begin{equation}
   ds^2=a(\eta)^2[-d\eta^2+\delta_{ij}dx^idx^j],\label{FRWmetric}
\end{equation}
where $\eta$ is the conformal time, $x^i$ are the 3 dimensional spatial Cartesian coordinates $(x,y,z)$, and $a(\eta)$ is the scale factor.

In order to preserve the FRW symmetries, the background scalar field is assumed to depend only on the conformal time:
\begin{equation}
    \vartheta = \theta(\eta).\label{FRWscalar}
\end{equation}

In the Palatini CS model, we also must assume an ansatz for the connection because it is an independent field. We follow the methodology presented in \cite{Tattersall2017}, in which a covariant approach to cosmology was proposed, based solely on the symmetries of the FRW background. According to this approach, we propose the following ansatz for the connection:
\begin{align}
    \Gamma^0_{00}& =f_1(\eta),\nonumber \\
    \Gamma^0_{ij}& =f_2(\eta) \delta_{ij},\nonumber \\
    \Gamma^i_{0j}& =\Gamma^i_{j0}=f_3(\eta)\delta^i_j, \label{FRWgamma}
\end{align}
and any other component vanishing. Here, $f_{1,2,3}$ are arbitrary functions of $\eta$ to be determined later by the equations of motion. 

Since for cosmology it is relevant to include the effect from additional matter components such as radiation or baryonic matter, we will also assume that there is an additional perfect fluid with the following stress-energy tensor:
\begin{equation}
    T^{mat}_{\mu \nu} =\left( \rho(\eta) + p(\eta)  \right) u_{\mu}u_{\nu} + p(\eta) g_{\mu \nu} ,\label{FRWfluid}
\end{equation}
where $\rho(\eta)$ is the energy density, $p(\eta)$ the pressure, and $u_{\mu}=(a,0,0,0)$ the 4-velocity of the fluid assumed to be at rest. The conservation of $T^{mat}_{\mu \nu}$ according to Eq.\ (\ref{Eq_mat_cons}) leads to:
\begin{equation}
    \rho'+3\mathcal{H}(\rho+p)=0,
\end{equation}
where primes denote derivatives with respect to the conformal time. This conservation equation together with the specification of an equation of state $p(\rho)$ completely characterize the background matter evolution.

Replacing the ansatz in Eqs.\ (\ref{FRWmetric})-(\ref{FRWfluid}) into the equations of motion obtained in Subsec.\ \ref{subsec_1CS_Eqns}, we obtain the background cosmological equations in the Palatini CS model. Using the equation for the connection we first find that
\begin{equation}
    f_1(\eta)=f_2(\eta)=f_3(\eta)=\mathcal{H}(\eta), \label{fi_sols}
\end{equation}
where $\mathcal{H}=a'/a$  the conformal Hubble parameter. We emphasize that this is the same result as if we had replaced the FRW metric into the Levi-Civita connection, as in the metric formalism.

From the metric equations, we obtain the Friedman equations:
\begin{align}
    6\kappa \mathcal{H}^2  & =  a^2\rho  + \beta  \left( \frac{1}{2}\theta^{'2}+ a^2  V(\theta)\right) ,\label{eq1} \\
   12\kappa \frac{ a^{''}}{a} & =-a^2(\rho - 3p) +\beta  \left(\theta^{'2} - 4 a^2 V(\theta)\right). \label{eq2}
\end{align}
In addition, we obtain a background equation for the scalar field:
\begin{equation}
    \theta^{''} + 2 \mathcal{H}\theta' + a^2\frac{dV(\theta)}{d \theta}  =0.
    \label{eq3}
\end{equation}
The specific evolution of the scalar field as a function of time is then determined by the choice of potential $V(\theta)$. In this paper, we do not dive into exploring this potential, and instead keep the system generic.

From these results, we see that $\alpha$ does not appear in the background equations, which means that the Chern-Simons interaction is irrelevant. This is because the  parity property of the CS interaction makes it incompatible with the
symmetries of FRW. In particular, for the CS interaction (and the whole action) to be parity invariant, $\vartheta$ must be parity odd, i.e.\ transform as $\hat{P}[\vartheta]=-\vartheta$ under the parity transformation $\hat{P}$ \cite{Alexander:2009tp}. However, on a cosmological background, $\vartheta$ does not depend on the spatial coordinates and hence does not satisfy the odd transformation required by the CS interaction. 
As a result, at the background level, both metric CS and Palatini CS theories give the same equations, which in turn are the same as those in GR with a minimally coupled scalar field.

The background equations for metric CS theory have been previously calculated in e.g.\ \cite{Guarrera:2007tu, PhysRevD.86.124031}, and agree with the results obtained in this section.

\subsection{Perturbations}
In order to describe the large scale inhomogeneities of the Universe, it is sufficient to consider linear perturbations around the FRW background for each field present in the action.

For the metric, we will have 
\begin{equation}
    g_{\mu \nu}=g_{(0) \mu \nu}(\eta) + \delta g_{\mu \nu}(\eta, \vec{x}),
\end{equation}
where $g_{(0)\: \mu \nu}$ is now the FRW metric in Eq.\ (\ref{FRWmetric}), and $\delta g_{\mu \nu}$ a small linear perturbation that can depend on space and time. For the scalar field, we will have:
\begin{equation}
    \vartheta =\theta(\eta) + \delta\vartheta(\eta,\vec{x}),  \label{PertScalar}
\end{equation}
with $\delta\vartheta(\eta,\vec{x})$ being the linear perturbation around the FRW solution $\theta(\eta)$.

It is known that, because of the symmetries of the background, it is convenient to decompose the metric perturbations into SVT: scalar, vector and tensor-type, depending on how they transform under spatial rotations \cite{dodelson2003modern, Mukhanov:2005sc}. In particular, the perturbation $\delta\vartheta$ is of scalar-type, while the 10 components of the metric are decomposed as: 
\begin{align}
ds^{2}&=  -a^2\left(1+2\Psi\right)d\eta^{2}+a^2\left(\partial_{i}B-S_{i}\right)d\eta dx^{i}\nonumber\\
& +a^{2}\left[\left(1-2\Phi\right)\delta_{ij}+2\partial_{i}\partial_{j}E+\partial_{i}F_{j}+\partial_{j}F_{i}+h_{ij}\right] dx^{i}dx^{j}\,,
\end{align}
where $\Psi(\eta,\vec{x})$, $\Phi(\eta,\vec{x})$, $E(\eta,\vec{x})$ and $B(\eta,\vec{x})$ are scalar-type perturbations; $F_i(\eta,\vec{x})$ and $S_i(\eta,\vec{x})$ are vector-type perturbations; $h_{ij}(\eta,\vec{x})$ are tensor-type perturbations. By construction, the vector and tensor perturbations are transverse, and the tensor perturbation is additionally traceless:
\begin{equation}
\partial^{i}F_{i}=\partial^{i}h_{ij}=h_{\phantom{i}i}^{i}=0\,, \label{TTh}
\end{equation}
where spatial indices are raised and lowered with $\delta_{ij}$.

In addition, there may be matter perturbations present as well. A perfect fluid will contain perturbations for its density, pressure and velocity. Then, there will be three scalar perturbations: $\delta \rho$, $\delta p$, and $\delta u$; and one transverse vector perturbation $\delta u_i$. Typical fluid models do not have tensor perturbations, unless they contain an anisotropic stress. From now on, we assume that matter does not contribute with tensor perturbations.

This SVT decomposition is useful because for linear perturbations around FRW, scalar, vector and tensor-type degrees of freedom decouple from each other, and can hence be studied separately. Scalar perturbations will directly impact matter inhomogeneities in the Universe, vector perturbations typically decay fast and become observationally irrelevant, while tensor perturbations describe the propagation of GWs.

In this paper, we will only analyze tensor perturbations, and leave scalar and vectors for the future. This means that we will ignore $\delta\vartheta$ and only keep the transverse-traceless metric perturbations $h_{ij}$:
\begin{equation}
    {ds^2} =a(\eta)^2[-d\eta^2+(\delta_{ij}+h_{ij}(\eta,\vec{x}))dx^idx^j].\label{metric_tensor}
\end{equation}
If we additionally assume, without loss of generality, that the GW is propagating along the $z$ axis, we can have a simple explicit expression for $h_{ij}$ satisfying (\ref{TTh}):
\begin{equation}
    h_{ij}(\eta,\vec{x})=\begin{pmatrix}
 h_{+}(\eta,z) & h_{\times}(\eta,z) & 0\\
 h_{\times}(\eta,z) & -h_{+}(\eta,z) & 0\\
 0 & 0 & 0
\end{pmatrix},
\label{h}
\end{equation}
where $h_+$ and $h_\times$ are the plus and cross polarization of GWs, respectively. When working with complex quantities, it will be useful to also introduce the alternative left and right-handed polarization basis:
\begin{align}
    h_R =\frac{h_{+}-ih_\times}{\sqrt{2}},\quad  h_L =\frac{h_{+}+ih_\times}{\sqrt{2}}.
     \label{rl}
\end{align}
As we will confirm later on, the GW polarizations $h_\times$ and $h_+$ will satisfy a set of coupled equations of motion when parity symmetry is broken. However, due to the translation and rotation invariance of the FRW background, as discussed in \cite{BeltranJimenez:2019xxx}, the $h_L$ and $h_R$ polarizations will satisfy decoupled equations, and hence they can be solved independently from each other.
In what follows, we calculate the equations of motion for $h_{L,R}$ in the metric and Palatini formalisms for Chern-Simons gravity. 

\subsubsection{CS: metric formulation}
Here, we re-derive the results on tensor perturbations for metric Chern-Simons gravity. We replace the perturbed metric (\ref{metric_tensor}) and scalar (\ref{PertScalar})---choosing $\delta\vartheta=0$---into the equations of motion obtained in Subsec.\ \ref{subsec_2ndorder}, and expand linearly in $h_{ij}$. 

There will be only two non-vanishing metric equations: 
\begin{widetext}
\begin{equation}
    \partial_{\eta}^2 h_R+ 2\mathcal{H} \partial_{\eta} h_R- \partial_{z}^2 h_R=  \displaystyle\frac{2 i \alpha(\theta^{''}\partial_{\eta}\partial_{z}h_R + \theta'(-\partial^3_{z}h_R+\partial^2_{\eta}\partial_{z}h_R))}{\kappa a^2},
    \label{gw1}
\end{equation}
\begin{equation}
    \partial_{\eta}^2 h_L  + 2\mathcal{H} \partial_{\eta} h_L - \partial_{z}^2 h_L=  - \displaystyle\frac{2 i \alpha (\theta^{''}\partial_{\eta}\partial_{z}h_L + \theta'(-\partial^3_{z}h_L+\partial^2_{\eta}\partial_{z}h_L ))}{\kappa a^2}.
    \label{gw2}
\end{equation}
\end{widetext}
We confirm that these equations are decoupled form each other and agree with previous results found in e.g.\ \cite{Alexander:2007kv, Yunes:2010yf, Yagi:2017zhb}. The LHS of these equations corresponds to the equations of motion for GW in GR, whereas the RHS  modifies the GW propagation due to the CS coupling. In this FRW background, these modifications contain third-order spatial derivatives but only second-order time derivatives, and thus avoid the Ostrogradski instability. Also, only two initial conditions will be required to fully specify the solutions, just like in GR.

While in GR, both $h_L$ and $h_R$ propagate in the same way, here we see that the two propagation equations differ by a sign on the RHS, which evidences the chiral-dependence introduced by the CS interaction. Also, notice that these equations only depend on $\beta$ and on the additional matter fluid indirectly through their effect on the scale factor $a$. 

Since we will compare to the Palatini formalism later, where the connection is an independent field, here we also explicitly show the non-vanishing terms of the perturbed Levi-Civita connection:
\begin{align}
    \delta\Gamma^{1}_{\;02}&=-\mathcal{H}h_{L}-\frac{1}{2}\partial_{\eta}h_{L}, & \delta\Gamma^{1}_{\;01}&=-\mathcal{H}h_{R}-\frac{1}{2}\partial_{\eta}h_{R}, \nonumber\\
    \delta\Gamma^{0}_{\;12}&=\frac{1}{2}\partial_{\eta}h_{L}, &\delta\Gamma^{0}_{\;11}&=\frac{1}{2}\partial_{\eta}h_{R},\nonumber\\
    \delta\Gamma^{2}_{\;31}&=-\frac{1}{2}\partial_{z}h_L, &\delta\Gamma^{1}_{\;31}&=-\frac{1}{2}\partial_{z}h_R,\nonumber\\
    \delta\Gamma^{3}_{\;21}&=\frac{1}{2}\partial_{z}h_L, &\delta\Gamma^{3}_{\;11}&=\frac{1}{2}\partial_{z}h_R,
    \
    \label{g1coef}
\end{align}
which do not have any explicit dependence on the Chern-Simons interaction.

\subsubsection{CS: Palatini formulation}
We now derive the tensor propagation equations in the Palatini CS model. In order to do this, we must start by complementing the metric and scalar perturbations with a connection perturbation:
\begin{equation}
    \Gamma^{\mu}_{\; \alpha  \beta} = \Gamma^{(0)\mu}_{\; \alpha  \beta}(\eta) +  \delta\Gamma^{\mu}_{\; \alpha  \beta}(\eta, \vec{x}),
\end{equation}
where $\Gamma^{(0)\mu}_{\; \alpha  \beta}$ is the FRW connection already described in Eqs.\ (\ref{FRWgamma}) and (\ref{fi_sols}), and $\delta\Gamma^{\mu}_{\; \alpha  \beta}$ is the linear perturbation. 

We again follow the methodology presented in \cite{Tattersall2017} and propose the following ansatz for the tensor-type perturbations in $\delta\Gamma^{\mu}_{\; \alpha  \beta}$:
\begin{align}
    \delta\Gamma^{\mu}_{\; \alpha  \beta}&=B_1(\eta) u^{\mu} \gamma_{1 \alpha \beta}+B_2(\eta) \gamma_{2 (\beta}^{\mu}u_{\alpha)}\nonumber\\
    & +B_3(\eta) \partial^{\mu}\gamma_{3 \alpha \beta} +B_4(\eta) \partial_{(\alpha}\gamma^{\mu}_{4 \beta)},\label{Gamma_pert_ansatz}
\end{align}
where $u_{\mu}=(a,0,0,0)$, $B_{1,2,3,4}$ are general functions that can depend on the cosmological background, and $\gamma_{1,2,3,4}$ are transverse-traceless linear perturbations with the same structure as \eqref{h}: each one contains a $(+,\times)$ polarization. This Ansatz is motivated by the fact that we are only interested in rank-2 tensor-type perturbations of $\Gamma^{\mu}_{\; \alpha  \beta}$ and all of these perturbations must have be accompanied by background coefficients that satisfy the FRW symmetries.
For simplicity, and without loss of generality, we will set $B_{1,2,3,4}=1$, which can be done by redefining appropriately the tensor perturbations $\gamma_{1,2,3,4}$.

Now that we have an appropriate perturbative ansatz for the connection, we proceed to replace (\ref{PertScalar}), (\ref{metric_tensor}) and (\ref{Gamma_pert_ansatz}) into the equations of motion obtained in Subsec.\ \ref{subsec_1CS_Eqns} and expand at linear order in the perturbations. 
Since the metric carries 1 tensor and the connection carries 4 tensors, we will initially obtain a total set of 10 coupled linear equations for all $+$ and $\times$ polarizations. In the right and left-handed polarization basis, we simplify the system to two decoupled sets of 5 equations for each circular polarization. The only difference between $L$ and $R$ polarization equations will be a sign in all the terms with $\alpha$, analogously to the equations for the metric CS model. It is thus convenient to introduce the general parameter:
\begin{align}
    \alpha_{R,L}=\alpha\cdot \varsigma_{R,L},
\end{align}
where $\varsigma_{R}\equiv 1$, $\varsigma_{L}\equiv -1$. Using $\alpha_{R,L}$, we then write both $L$ and $R$ polarization equations in a compact form. From the connection equations we obtain:
\begin{widetext}
\begin{align}
     a^2\kappa  (\partial_{\eta}h_{R,L}-(a^2 \gamma _{2\text{  }R,L}+  \partial_{\eta}\gamma _{4\text{  }R,L}{}))
   &=i\alpha_{R,L}  \theta '(a^2  \partial_z\gamma _{2\text{}R,L}{}+\partial_{\eta}\partial_{z}\gamma_{4\text{  }R,L}{}), \label{gamma_eq1}\\
   a^2\kappa (-\partial_{z}\gamma _{4\text{
 }R,L}{}+\partial_{z}h_{R,L}{})&=- i\alpha_{R,L} \theta'(\mathcal{H}( a^2  \gamma _{2\text{  }R,L}
    + \partial_{\eta}\gamma _{4\text{  }R,L}{})-  \partial^{2}_{z}\gamma _{4\text{  }R,L}{}),\\
    a^2 \kappa ((2a^2 \gamma _{1\text{  }R,L}+2\partial_{\eta}\gamma _{3\text{  }R,L}{}+a^2\gamma _{2\text{  }R,L}+ \partial_{\eta}\gamma _{4\text{  }R,L}{})&+2\mathcal{H} (2 \gamma _{3\text{  }R,L}+h_{R,L}))  \nonumber\\
    & = -  i \alpha_{R,L}\theta'\left(2( a^2 \partial_z\gamma _{1\text{  }R,L}{}+\partial_{\eta}\partial_z\gamma _{3\text{}R,L}{})+ \mathcal{H} (2 \partial_z\gamma _{3\text{  }R,L}{}
    +\partial_z\gamma _{4\text{  }R,L}{})\right),\\
   a^2 \kappa (2 \partial_z\gamma_{3\text{  }R,L}{}+\partial_z\gamma _{4\text{  }R,L}{})        & =-i\alpha_{R,L} \theta' \left(2\partial^2_z\gamma _{3\text{  }R,L}{}+\mathcal{H}(a^2  \gamma _{2\text{  }R,L} +   \partial_{\eta}\gamma _{4\text{  }R,L}{})\right).
    \label{eqlc}
\end{align}
From the metric equations we obtain:
\begin{equation}
    4a\mathcal{H}(a \gamma _{1\text{  }R,L}+\partial_{\eta}\gamma _{3\text{  }R,L}{})+a\mathcal{H}(a^2\gamma _{2\text{  }R,L}+\partial_{\eta}\gamma _{4\text{}R,L}{} ) +a(a^2 \partial_{\eta}\gamma _{1\text{  }R,L}{}+\partial^2_{\eta}\gamma _{3\text{  }R,L}{})
    + a^{''}(2 \gamma _{3\text{}R,L}+h_{R,L} )
    -a \partial^2_z\gamma _{3\text{  }R,L}{}=0.
    \label{eqlm}
\end{equation}
\end{widetext}
We emphasize again that these equations only contain first-order derivatives of the connection and metric. The only reason why we see second-order derivatives of the tensors $\gamma_{2}$ and $\gamma_{4}$ is because their ansatz in Eq.\ (\ref{Gamma_pert_ansatz}) contained one derivative. We also see explicitly again that the Chern-Simons interaction does not appear at all in the metric equation (\ref{eqlm}) and instead we see the parameter $\alpha$ only present in the connection equations.

Next, we proceed to manipulate these equations in order to obtain a single equation for $h_{L,R}$, if possible. Because the left and right-handed polarizations are decoupled, we can focus only on one circular polarization and recover the other one replacing $\alpha \xrightarrow[]{} - \alpha$. We start by using  the  Fourier transform to go to momentum space:
\begin{align}
    \gamma_{a\; R,L}(\eta,z) & =\int d^3k\; \bar{\gamma}_{a\; R,L}(\eta,\vec{k})\; e^{i\vec{k}\cdot \vec{x}},\nonumber\\
    h_{R,L}(\eta,z) & =\int d^3k\; \bar{h}_{R,L}(\eta,\vec{k})\; e^{i\vec{k}\cdot \vec{x}},
    \label{fourier}
\end{align}
where $a=\{1,2,3,4\}$. We will continue assuming that the wave propagates along $z$, that is, $\vec{k}=(0,0,k)$. Importantly, different $k$ modes will propagate independently at the linear perturbation level due to the translation symmetry of the FRW background.

We first notice that in the set of equations (\ref{gamma_eq1})-(\ref{eqlc}), $\gamma_{1,2}$ always appear in the combinations $\tilde{\gamma}_{1} = \gamma_{1}a^2+\partial_\eta\gamma_{3}$ and $\tilde{\gamma}_{2} =\gamma_{2}a^2+\partial_\eta\gamma_{4}$ such that $\tilde{\gamma}_{1,2}$ absorb all the derivatives of $\gamma_{3,4}$. As a result, we use (\ref{gamma_eq1})-(\ref{eqlc}) to solve algebraically for $\tilde{\gamma}_{1,2}$ and $\gamma_{3,4}$ in terms of $h$ and $h'$,
which result into the following schematic form for the connection perturbation functions: 
\begin{align}
    \gamma_{1\; R,L}&=b_1 h_{R,L}+ c_1 h'_{R,L}+ d_1 h''_{R,L},\nonumber\\
    \gamma_{2\; R,L}&=b_2 h_{R,L}+ c_2 h'_{R,L}+d_2h''_{R,L},\nonumber\\
    \gamma_{3\; R,L}&=b_3 h_{R,L}+ c_3 h'_{R,L},\nonumber\\
    \gamma_{4\; R,L}&=b_4 h_{R,L}+ c_4 h'_{R,L}, \label{gammas_fc_h}
\end{align}
where $\{b_a,c_a,d_a\}$ are coefficients that depend on the background functions.  After replacing the expressions found in (\ref{gammas_fc_h}) into Eq.\ (\ref{eqlm}) we finally obtain a final differential equation for the metric perturbations, whose schematic form is as follows:
\begin{align}
    \bar{h}_{R,L}^{''} &+ \left(\frac{2\mathcal{H}+  b_{nm} \alpha_{R,L}^n k^m }{1+f_{nm} \alpha_{R,L}^nk^m}\right) \bar{h}'_{R,L}\nonumber\\
    &+ \left(\frac{1 +c_{nm}\alpha_{R,L}^nk^m}{1+f_{nm} \alpha_{R,L}^nk^m} \right)k^2\bar{h}_{R,L}=0 ,
    \label{heq_1CS}
\end{align}
where the coefficients $b_{nm}$, $c_{nm}$ and $f_{nm}$ are given in Appendix \ref{app:h_coeffs} and are sole functions of time, via the scale factor $a$ and the background scalar field $\theta$. Here, we have an implicit sum over $n$ and $m$ in numerators and denominators, such that $n>0$ denotes the power in which $\alpha$ appears, and $m\geq 0$ the power of $k$. We find that the highest powers of $\alpha$ appear as $b_{42}\alpha^4k^2$, $c_{53}\alpha_{L,R}^5k^3$ and $f_{42}\alpha^4k^2$. We generically also find $m\leq n$. 
Note that Eq.\ (\ref{heq_1CS}) is written such that it is manifest that we recover the propagation equation of motion in GR whenever $\alpha \rightarrow 0$.

Notice that the initial metric equation (\ref{eqlm}) contained only first-order time derivatives of the metric and hence required only one metric initial condition. However, after eliminating the Christoffel perturbations $\gamma_{1,2,3,4}$, we obtain a second-order derivative equation for $h$ in (\ref{heq_1CS}), which now requires two initial conditions to fix entirely the metric solution, similarly to GR. This means that Palatini CS gravity propagates the same number of helicity-2 polarizations as GR and other well-known modified gravity models such as $f(R)$ theories, on a cosmological background \cite{Nojiri_2011}. Nonetheless, Palatini CS predicts a different propagation equation for the two helicity-2 polarizations due to parity violation, contrary to GR and $f(R)$ theories which predict the same propagation equation for left and right-handed polarizations. Also note that, similarly to $f(R)$ theories, we do expect Palatini CS to propagate at least one additional helicity-0 polarization associated to the scalar field perturbation $\delta\theta$. The consequence of this and other helicity polarizations will be left for a future study. 

Note that the second-order time derivatives of $h$ in (\ref{heq_1CS}) appear because Eq.\ (\ref{eqlm}) contained first-order time derivatives of $\gamma_1$ and second-order time derivatives of $\gamma_3$ (while the first-order time derivative of $\gamma_4$ does not contribute with $h^{''}$ terms since it again appears in the combination $\tilde{\gamma}_2$). 

As we did in \eqref{g1coef}, we are interested in the perturbed connection coefficients. In order to compare the results with the ones obtained in the previous section we will make an expansion around $\alpha=0$ (valid when $\alpha \theta' k/\kappa\ll 1$) to quadratic order, to obtain simpler expressions. The non-zero coefficients in this Palatini formalism are given by:
\begin{widetext}
\begin{align}
    \delta\Gamma^{1}_{\;02}&\approx -\mathcal{H}\bar{h}_{L}-\frac{1}{2}\partial_{\eta}\bar{h}_{L}-\displaystyle\frac{i\alpha\theta'k\partial_{\eta}\bar{h}_{R,L}}{a^2\kappa}+\mathcal{O}(\alpha^2),
    &\delta\Gamma^{1}_{\;01}&\approx -\mathcal{H}\bar{h}_{R}-\frac{1}{2}\partial_{\eta}\bar{h}_{R}+\displaystyle\frac{i\alpha\theta'k\partial_{\eta}\bar{h}_{R}}{a^2\kappa}+\mathcal{O}(\alpha^2),\nonumber\\
    \delta\Gamma^{0}_{\;12}&\approx\displaystyle\frac{1}{2}\partial_{\eta}\bar{h}_{L}- \displaystyle\frac{i\alpha\theta'k\partial_{\eta}\bar{h}_{L} }{2a^2\kappa}+\mathcal{O}(\alpha^2), &\delta\Gamma^{0}_{\;11}&\approx\displaystyle\frac{1}{2}\partial_{\eta}\bar{h}_{R}+ \displaystyle\frac{i\alpha\theta'k\partial_{\eta}\bar{h}_{R} }{2a^2\kappa}+\mathcal{O}(\alpha^2),\nonumber\\
    \delta\Gamma^{2}_{\;31}&\approx -\displaystyle\frac{1}{2}\partial_{z}\bar{h}_{L}+\displaystyle\frac{i\alpha\theta'( k^2\bar{h}_{L}-\mathcal{H}\partial_{\eta}\bar{h}_{L})}{a^2\kappa}+\mathcal{O}(\alpha^2), &\delta\Gamma^{1}_{\;31}&\approx-\displaystyle\frac{1}{2}\partial_{z}\bar{h}_{R}-\displaystyle\frac{i\alpha\theta'( k^2\bar{h}_{R}-\mathcal{H}\partial_{\eta}\bar{h}_{R})}{a^2\kappa}+\mathcal{O}(\alpha^2),\nonumber\\
    \delta\Gamma^{3}_{\;21}&\approx\displaystyle\frac{1}{2}\partial_{z}\bar{h}_{L}-\displaystyle\frac{i\alpha\theta'( k^2\bar{h}_{L}-\mathcal{H}\partial_{\eta}\bar{h}_{L})}{a^2\kappa}+\mathcal{O}(\alpha^2), & \delta\Gamma^{3}_{\;11}&\approx\displaystyle\frac{1}{2}\partial_{z}\bar{h}_{R}+\displaystyle\frac{i\alpha\theta'( k^2\bar{h}_{R}-\mathcal{H}\partial_{\eta}\bar{h}_{R})}{a^2\kappa}+\mathcal{O}(\alpha^2).
\end{align}

\end{widetext}
We can see that the procedure is consistent because if we set $\alpha=0$, i.e.\ the Chern-Simons term does not contribute, we recover the linearly perturbed Levi-Civita connection, in agreement with Eq.\ (\ref{g1coef}). In addition we see that the perturbed connection is clearly different to the metric formalism, with linear and higher-order $\alpha$ terms appearing. Here, the connection does not depend solely on the metric, but it also depends on how the background scalar field $\theta$ evolves with cosmological time.


\section{Gravitational Waves}\label{sec:GWs}
Chern-Simons has been analyzed in the context of both early and late-time cosmology. During early times, the scalar $\vartheta$ can describe the inflationary field or some additional primordial field \cite{Lue:1998mq, Alexander:2004us, Alexander:2006lty, Alexander:2011hz, Satoh:2007gn, Cai:2016ihp, Odintsov:2022hxu}, in which case it is relevant to determine the statistical initial conditions of the inhomogeneities in the Universe. In that context, previous authors have analyzed the behavior of GWs and the prediction for its power spectrum, which determines the GWs generated during the early Universe. These chiral GWs can also affect higher-order correlation functions of primordial matter perturbations.
During late times, the scalar field may describe a new component, such as dark matter \cite{Yoshida:2017cjl,Nojiri_2011, Nojiri_2019, Nojiri:2020pqr, Jung:2020aem, Tsutsui:2022zos} or dark energy. In the context of GWs, this late-time scalar field can affect how GWs from astrophysical sources propagate towards us, which has been recently constrained with current binary black hole GWs in \cite{Okounkova:2021xjv}.

In this section, we will discuss the phenomenology of the propagation of GWs in Palatini CS, and compare to metric CS gravity.
In both formalisms we find the following schematic linear equation for $L$ and $R$ polarizations in momentum space: 
\begin{equation}
    \bar{h}_{R,L}^{''}(\eta,k)+2\Xi_{R,L}(\eta,k) \bar{h}'_{R,L}(\eta,k)+ \omega^2_{R,L}(\eta,k)\bar{h}_{R,L}(\eta,k)=0 ,
    \label{h_eq}
\end{equation}
where the coefficients $\Xi_{R,L}(\eta,k)$ and $\omega_{R,L}(\eta,k)$ generically depend on time due to the cosmological expansion and momentum $k$. This equation is analogous to a damped harmonic oscillator, with natural frequency $\omega$ and friction  $\Xi$.

If we assume that $\Xi_{R,L}$ and $\omega_{R,L}$ evolve on cosmological timescales, while the period of the GW is much shorter, i.e.\ $\mathcal{H}\ll k$, we can solve \eqref{h_eq} using the WKB approach \cite{Dingle1968}. Under this approximation, the leading-order solution is given by
\begin{equation}
    \bar{h}_{R,L}(\eta,k)\approx A^{\pm}_{R,L} e^{-\int \Xi_{R,L} d\eta}e^{\pm i\int \Omega_{R,L} d\eta},\label{eq_hsol}
\end{equation}
where we have introduced arbitrary proportionality constants $A^{\pm}_{R,L}$ that depend on the initial conditions, and we have introduced the net oscillation frequency $\Omega^2\equiv \omega^2-\Xi^2$. Note that for each polarization $L$ or $R$, there will be two independent solutions to \eqref{h_eq}, which correspond to the two possible signs in the complex exponential in Eq.\ (\ref{eq_hsol}), which describe waves propagating along $z$ in opposite directions. As a comparison, in GR $\Xi_{R}=\Xi_L=\mathcal{H}$ and $\omega_{R}=\omega_{L}=k$, so that both polarizations propagate in the same way and their evolution only depends on the cosmological expansion history  through $\mathcal{H}$. The modifications to GR come in Eq.\ (\ref{eq_hsol}) via the integration of $\Xi_{R,L}-\mathcal{H}$ and $\omega_{R,L}-k$ from the time of emission to the time of detection.

From Eq.\ (\ref{eq_hsol}) we see that the L and R polarizations can suffer an amplitude change and/or a phase change. When $\Xi_R\not=\Xi_L$ we say that there is amplitude birefringence since the amplitude evolution of the GWs depends on the polarization. In this case, the polarization can change e.g.\ from purely linear (when $|h_L|=|h_R|$) to elliptical $|h_L|\not=|h_R|$. For this reason, we say that amplitude birefringence changes the ellipticity of the GW polarization.

\begin{figure}[h!]
	\includegraphics[width = 0.25\textwidth]{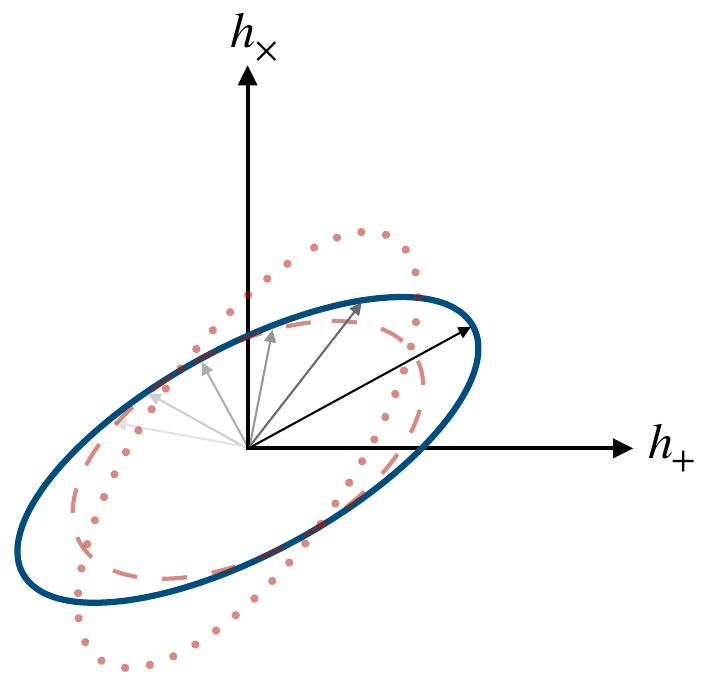}
	\caption{For a basis of two linear GW polarizations, $h_+$ and $h_\times$, any GW polarization can be described as a vector in the ($h_+$, $h_\times$) plane. This diagram illustrates how a GW polarization changes in time (black and gray vectors show a polarization that rotates counter-clockwise as time evolves) when it has a given elliptical polarization (blue contour). The red dotted and red dashed contours show examples of polarizations with a different polarization plane (but same ellipticity) and different ellipticity (but same polarization plane), respectively.}
	\label{fig:polprops}
\end{figure}

When $\omega_R\not=\omega_L$ we say that there is velocity birefringence since the dispersion relation (which determines the phase evolution and the velocity propagation of the signal) depends on the polarization. In particular, $\omega_{R,L}(k)$ is the dispersion relation of the GWs\footnote{Note that technically the dispersion relation of the GWs will be set by $\Omega(k)$ instead of $\omega(k)$. However, in the approximation that $k/\mathcal{H}\gg 1$ and when all the modifications of gravity describe small corrections from GR, then $\Omega\approx \omega$.}, from which we can obtain the propagation velocity as the group velocity of the waves $v_{R,L}=\partial\omega_{R,L}/\partial k$. 
For a monochromatic wave (i.e.\ fixed value of $k$), a polarization-dependent velocity will lead to a phase shift between $h_L$ and $h_R$ that can be interpreted as a rotation of the polarization plane. See Fig.\ \ref{fig:polprops} for a toy illustration of amplitude and velocity birefringence (see also a review on GW polarization in \cite{Isi:2022mbx}). See also in \cite{Zhao:2019xmm} a compilation of other parity-breaking gravity models that exhibit amplitude and velocity birefringence, and their effect on GWs. In GR, GWs do not change their polarizations during cosmological propagation, and hence do not suffer from amplitude nor velocity birefringence. 

If we had a wavepacket, with a wide range of wavelengths, and the propagation velocity depended on $k$, then the L and R polarizations would have a dispersive velocity that would induce phase distortions in the GW signal during propagation
\cite{Will:1997bb, Mirshekari:2011yq,Mastrogiovanni:2020gua, LISACosmologyWorkingGroup:2022wjo, Ezquiaga:2022nak}. Similarly, if $\Xi$ depended on $k$, it would lead to amplitude distortions during propagation. In the case of GWs from binary systems, whose phase and amplitude evolution is predicted by GR, these distortions can be observed and used to constrain parity-breaking modified gravity theories.

As previously discussed, the Chern-Simons interaction in the metric formalism contains higher derivative terms, and thus this theory must be considered as an EFT where the CS corrections to GR are small. For this reason, in this section we will make a small $\alpha$ approximation: $\alpha k\theta'/\kappa \ll 1$\footnote{We also assume that $\theta$ evolves on cosmological timescales such that $\theta^{''}\sim \theta'\mathcal{H}$.}. In metric CS, the GW equation of motion (\ref{h_eq}) has the following coefficients:
\begin{align}
    \omega_{R,L}^2&=k^2, \label{w_2CS}\\
    2\Xi_{R,L}&\approx  2\mathcal{H}-\frac{2  (\theta^{''}-2\theta'\mathcal{H})k\alpha_{R,L}}{\kappa a^2}  \nonumber\\
    &-\frac{4\theta' (\theta^{''}-2\theta^{'}\mathcal{H})(k \alpha)^2}{ \kappa^2 a^4}+\mathcal{O}\left(
(k\alpha)^3\right) ,\label{Xi_2CS}
\end{align}
where the expression in Eq.\ (\ref{w_2CS}) is exact (for any $\alpha$), whereas in Eq.\ (\ref{Xi_2CS}) we have made the small $\alpha$ approximation (the general expression for any $\alpha$ is rather simple and is presented explicitly in Appendix \ref{app:h_coeffs}). Here, even powers in $\alpha$ will not break parity symmetry, whereas odd powers in $\alpha$ will break parity. We thus confirm previous results on the fact that CS induces amplitude birefringence but no velocity birefringence. This means that during propagation, the ratio $|\bar{h}_L/\bar{h}_R|$ will suffer modifications with respect to GR, but the relative phase between $\bar{h}_L$ and $\bar{h}_R$ will be the same as in GR. In particular, depending on the sign of the second term in the RHS of Eq.\ (\ref{Xi_2CS}), one polarization will grow exponentially in time while the other one will decay exponentially, according to Eq.\ (\ref{eq_hsol}). 

We emphasize that while previous analyses have focused on making a linear expansion in $\alpha$, we see in Eq.\ (\ref{Xi_2CS}) that there are higher-order terms but one expects them to be suppressed.  Nevertheless, the third term on the RHS of Eq.\ (\ref{Xi_2CS}) is still expected to bring modifications to GR by introducing a frequency-dependent amplitude modulation of the GW signal.

Next, we discuss the results on Palatini Chern-Simons. While in this theory there is no need to impose a priori that the CS interaction is small, we will do so motivated by current observations from binary mergers, which have been shown to be in agreement with GR so far \cite{LIGOScientific:2021sio}. In this formalism, we obtain the following $\omega$ coefficient for the GW propagation equation:
\begin{align}
    \omega_{R,L}^2&\approx k^2-\displaystyle\frac{3\theta'(2a\theta^{''}\mathcal{H}+a^{''}\theta'-4a\mathcal{H}^2\theta')(k \alpha)^2}{a^5 \kappa^2} \nonumber\\
    &+ \mathcal{O}\left((k\alpha)^3\right).
    \label{w_1CS}
\end{align}
From Eq.\ (\ref{w_1CS}) we see that the dispersion relation of GWs is modified with respect to GR, and the leading correction goes as $\alpha^2$. This means that this correction does not violate parity and, in this case, it also goes as $k^2$ which means that it introduces a time-dependent (yet frequency-independent) correction to the propagation speed of GWs\footnote{Recall that the propagation speed is determined by the group velocity, calculated as $\partial\omega/\partial k$ so that a $k^2$ correction in $\omega^2$ leads to a frequency-independent change in the GW group velocity.}. Observations from cosmic rays \cite{Moore:2001bv} and from the binary neutron star merger GW170817 \cite{LIGOScientific:2017vwq} and its electromagnetic counterparts \cite{Goldstein:2017mmi, Savchenko:2017ffs} set tight bounds on the propagation speed $c_T$ of GWs compared to the speed of light today: $|c_T-1|<10^{-15}$ \cite{LIGOScientific:2017zic}. 
If we consider Palatini CS as a late-time cosmological modification of gravity, we can calculate the GW speed from Eq.\ (\ref{w_1CS}) and impose the following constraint:
\begin{equation}
   |c_T-1|\approx 3H_0|\theta'_{0}(\theta^{''}_0-H_0\theta'_{0})|\left(\frac{\alpha }{\kappa}\right)^2 \lesssim 10^{-15} , \label{eq_cT}
\end{equation}
where the subscript $0$ refers to the value today, and we have assumed a dark-energy dominated Universe. From (\ref{eq_cT}) we thus obtain a constraint on the combined behavior of $\alpha$ and $\theta$ during late times \footnote{Note that Palatini CS predicts a time-varying $c_T(\eta)$ and observations constrain the averaged velocity over a time window between emission to detection. Since the source for GW170817 was really close (with redshift $z\approx 0.01$), in (\ref{eq_cT}) we assume $c_T$ to be effectively constant at its value today.}. This result means that linear effects would have to be smaller than $|\alpha \theta_0'H_0/\kappa| \lesssim \mathcal{O}(10^{-7})$, unless a special cancellation was causing $c_T$ to be so close to 1. 

Note that the next-to-leading order correction $(k\alpha)^3$ in Eq.\ (\ref{w_1CS}) will bring phase distortions that are distinct from GR and can also be constrained with LIGO/Virgo data \cite{Wang:2020cub, Zhao:2022pun}. From (\ref{eq_cT}), those effects could still be of order $10^{-21}(k/H_0)$, which can be measurable with current and planned GW detectors since ground-based GW detectors are sensitive to $k\sim 10-10^3$Hz and thus $k/H_0\sim 10^{19}-10^{21}$. 
The previous analyses \cite{Wang:2020cub, Zhao:2022pun} assumed a specific simple time evolution for these $k^3$ correction terms, that may not correspond to that predicted in Palatini CS depending on how $\theta$ evolves. We will leave a detailed quantitative analysis on these higher-order effects for the future, but since the phase evolution of GWs is measured to great precision, these effects have the potential to give one of the tightest constraints on Palatini CS. Indeed, if we were to take the results of \cite{Zhao:2022pun} at face value (ignoring time evolution assumptions), one would obtain an estimate of $|\alpha \theta_0'H_0/\kappa|^3 \lesssim \mathcal{O}(10^{-42})$\footnote{This has been estimated taking their parameter $\zeta < 10^{-16}$m \cite{Zhao:2022pun}  and making it dimensionless by calculating $\zeta H_0$.}. This means that the linear effects would be, at most, of order  $|\alpha \theta_0'H_0/\kappa| \lesssim \mathcal{O}(10^{-14})$, which is much more constraining that the GW propagation speed result in Eq.\ (\ref{eq_cT}). This result emphasizes the fact that a purely linear expansion in $\alpha$ may not always be the appropriate truncation order since higher-order smaller terms may still be easier to observe and provide crucial about the theory.

On the other hand, for $\Xi$ we find the following expression in Palatini CS:
\begin{widetext}
\begin{equation}
    2\Xi_{R,L}\approx  2\mathcal{H}-\frac{2 (\theta^{''}-2\theta'\mathcal{H})k\alpha_{R,L}}{\kappa a^2}-\frac{2 \left(3 a^{3}\mathcal{H}^2\theta'\theta^{''} +3 a^2 \mathcal{H}a^{''} \theta^{'2}-9 a^{3}\mathcal{H}^3\theta^{'2}-a^3\theta' \theta^{''} k^2+2 a^3 \mathcal{H}\theta^{'2} k^2\right)\alpha^2}{a^7 \kappa^2 } + \mathcal{O}((k\alpha)^3). \label{Xi1st}
\end{equation}
\end{widetext}
From here we see that the leading-order correction is exactly the same as in the metric CS formalism (c.f.\ Eq.\ (\ref{Xi_2CS})). The expressions now differ in the next-to-leading correction with $\alpha^2$.  

When the deviations from GR are assumed to be small and the linear $\alpha$ term is dominant, it is then possible to translate current GW constraints for metric CS gravity \cite{Okounkova:2021xjv} (which truncate the $\alpha$ expansion to linear order) directly onto constraints for Palatini CS. These constraints roughly give\footnote{The constraint in (\ref{AmpBire_constraint}) is technically valid only when $(\theta^{''}-2\theta'\mathcal{H})$ does not vary with conformal time, since that is the assumption made in \cite{Okounkova:2021xjv}. For other time evolutions the constraints may vary since current detected GW sources are present at redshift up to $z\sim 1$ and may have considerably time variations.}:
\begin{equation}
    |(\theta^{''}_0-2\theta'_0H_0)|\left(\frac{k}{H_0}\right)\left(\frac{\alpha}{\kappa}\right)\lesssim \mathcal{O}(1) \label{AmpBire_constraint},
\end{equation}
but since $k/H_0\sim 10^{19}-10^{21}$ for the LIGO frequency sensitivity range, this means that the linear effects are of order $|\alpha \theta_0'H_0/\kappa|\lesssim \mathcal{O}(10^{-20})$, which is much tighter than the velocity and cubic phase distortions previously discussed.
Note that amplitude birefringence is a frequency-dependent effect that distorts a binary waveform, due to the $k$ dependence of $\Xi_{R,L}$ in Eq.\ (\ref{Xi1st}). In the work of  \cite{Okounkova:2021xjv}, the authors ignored this $k$ dependence for simplicity, and obtained a constraint that could be interpreted as valid for $k\sim 100$Hz, which is the frequency of maximum sensitivity for LIGO. A future analysis including the frequency dependence of amplitude birefringence will have to be performed, which is expected to improve the results of \cite{Okounkova:2021xjv} by a few orders of magnitude. In addition, forecasts on CS constraints due to amplitude birefringence for future GW detectors such as LISA have been performed in \cite{Alexander:2007kv, Hu:2020rub}, which show that current constraints could be further improved by a few more orders of magnitude.

In summary, Palatini CS makes novel predictions that induce velocity modifications to GR, contrary to metric GR. Nevertheless, we have found that these new effects may yield weaker constraints than those of amplitude birefringence, due to the fact that they appear as higher-order corrections in the Chern-Simons coupling.
This means that in practice metric and Palatini CS will behave similarly when observing the propagation of astrophysical GWs.
Nevertheless, a future comprehensive analysis on  scalar cosmological perturbations will confirm whether Palatini CS gravity predicts any other late-time features that could be distinct from metric CS and falsifiable with galaxy surveys or cosmic microwave background observations.

\section{Discussion}\label{sec:discussion}
In this paper we have studied Chern-Simons (CS) modified gravity using the Palatini approach---dubbed Palatini Chern-Simons gravity. Here, the  metric and the connection are considered to be independent dynamical fields. We obtain the full nonlinear equations of motion and confirm that they contain only up to first-order derivatives of the metric and connection, and hence avoid the instabilities  that appear in the usual metric CS formalism due to higher-derivative interactions. These equations determine in a dynamical way the relationship between the metric and the connection, and we find that the connection generally differs from the Levi-Civita connection.

In order to illustrate what new features  Palatini CS exhibits compared to metric Chern-Simons, we analyze the cosmological evolution of the Universe, focusing on the background expansion history and on the propagation of gravitational waves. While at the background level, both theories make the same predictions, we find that they differ at the level of cosmological perturbations. 

We find that, contrary to the metric CS model, Palatini CS introduces the effect of GW velocity birefringence, in which  the dispersion relation of GWs is modified with respect to GR, and is generically different for the two GW polarizations. Nevertheless, we show that for a small CS coupling parameter, there is an overall shift in the GW propagation speed $c_T$, regardless of the polarization. By  considering Palatini CS as a modified gravity theory affecting the late-time cosmological evolution of the Universe, we discuss observational bounds on $c_T$ and obtain initially weak constraint on Palatini CS. We also discuss higher-order corrections to GR that break parity and induce phase distortions of the waveform, such that they are easier to observe and lead to much tighter constraints than those from $c_T$.

Furthermore, we show that, similarly to metric Chern-Simons, Palatini Chern-Simons predicts GW amplitude birefringence, a phenomenon where the amplitude evolution of the GWs depends on their polarization. While the specific way in which amplitude birefringence happens in metric and Palatini CS theories is generally different, they do coincide in the limit in which  the modifications to General Relativity are small. Thinking again of Palatini CS as late-time cosmological model, we thus apply current metric CS constraints directly to Palatini CS. The bound is found to be much tighter than those coming from velocity birefringence, and are expected to improve in the future. This highlights the fact that Palatini CS is expected to behave similarly to metric CS when observing the propagation of astrophysical GWs. 

Given the constraints on late-time Palatini CS, in the future it will be interesting to analyze this gravity theory in different contexts. For example, metric Chern-Simons has been widely studied as a possible inflationary theory, in which case the power spectrum of primordial matter and GWs is modified with respect to canonical parity-preserving inflationary models, and could be constrained using observations from galaxy surveys and the cosmic microwave background. The cosmological calculations performed in this paper could be then extended to calculate the primordial power spectrum and analyze the Palatini CS predictions.

Finally, it would also be interesting to analyze Palatini CS as modifications to GR in inhomogeneous environments. For instance, tests on metric CS have been performed, using Solar System \cite{Smith:2007jm} and binary pulsar \cite{Yunes:2008ua,Ali-Haimoud:2011wpu} observations. It is possible to use these observations to constrain Palatini CS, which would require to analyze its predictions on spherically symmetric backgrounds. Relatedly, metric CS has also been shown to produce a modification on the emitted GW from compact objects, using perturbative approaches \cite{Pani:2011xj, Canizares:2012is, Yagi:2012vf, Loutrel:2022tbk} as well as nonlinear numerical simulations \cite{Okounkova:2019dfo}. The same analyses could now be performed with Palatini CS to obtain its predictions. Since Palatini CS is already a first-order derivative theory, it will not require any additional approximation to be solved numerically, contrary to metric CS gravity.

\section{ACKNOWLEDGMENTS}
F.\ S.\ thanks Felipe Canales for fruitful discussions and checking some of the calculations.
M.\ L.\ was supported by the Innovative Theory Cosmology fellowship at Columbia University. M.\ B.\ was partially supported by Fondecyt Grant \#1201145 (Chile).

\appendix

\section{GW Equation}\label{app:h_coeffs}
We write here all of the coefficients $b_{nm}$, $c_{nm}$ and $f_{nm}$ that determine the propagation equation of GWs in the Palatini CS model, according to Eq.\ (\ref{heq_1CS}). These coefficients are expressed solely in terms of the scale factor $a$ and the background scalar field $\theta$.

The $b_{nm}$ coefficients are:
\begin{align}
    b_{11}&=\frac{2\theta''-10\mathcal{H}\theta'}{a^2\kappa},\\
    b_{20}&=\frac{12a\mathcal{H}^3\theta'^2-6a\mathcal{H}^2\theta'\theta''-6\mathcal{H}a''\theta'^2}{a^5\kappa^2},\\
    b_{22}&=\frac{14\mathcal{H}\theta'^2-4\theta'\theta''}{a^4\kappa^2},\\
    b_{33}&=\frac{2\theta'^2\theta''-6\mathcal{H}\theta'^3}{a^6\kappa^3},\\
    b_{42}&=\frac{2a\mathcal{H}^2\theta'^3\theta''-4\mathcal{H}a''\theta'^4-4a\mathcal{H}^3\theta'^4}{a^{9}\kappa^4}.
\end{align}
The $c_{nm}$ coefficients are: 
\begin{align}
    c_{11}&=\frac{3\theta'}{a^2\kappa},\\
    c_{22}&=\frac{3\theta'^2}{a^4\kappa^2},\\
    c_{20}&=\frac{9a\mathcal{H}^2\theta'^2-6a\mathcal{H}\theta'\theta''-3a''\theta'^2}{a^5\kappa^2},\\
    c_{33}&=\frac{-\theta'^3}{a^6 \kappa^3},\\
    c_{31}&=\frac{9a\mathcal{H}\theta'^2\theta''+8a''\theta'^3-10a^2\mathcal{H}^2\theta'^3}{a^7\kappa^3},\\
    c_{42}&=\frac{-(a\mathcal{H}^2\theta'^4+3a\mathcal{H}\theta'^3\theta''+7a''\theta'^4)}{a^{9}\kappa^4},\\
    c_{53}&=\frac{2a''\theta'^5+2a\mathcal{H}^2\theta'^5}{a^{11}\kappa^5}.
\end{align}
The $f_{nm}$ coefficients are:
\begin{align}
    f_{11}&=\frac{3\theta'}{a^2\kappa},\\
    f_{20}&=\frac{3\mathcal{H}^2\theta'^2}{a^4\kappa^2},\\
    f_{22}&=\frac{3\theta'^2}{a^4\kappa^2},\\
    f_{31}&=\frac{5\mathcal{H}^2\theta'^3}{a^6\kappa^3},\\
    f_{33}&=\frac{\theta'^3}{a^6\kappa^3},\\
    f_{42}&=\frac{2\mathcal{H}^2\theta'^4}{a^{8}\kappa^4}.
\end{align}

Next, we also write the explicit expression for the coefficient  $\Xi_{L,R}$ in Eq.\ (\ref{h_eq}) for the metric CS model:
\begin{equation}
    \Xi_{L,R} = \frac{\alpha_{R,L}\theta'' k+a^2\kappa\mathcal{H}}{2\alpha_{R,L}\theta'k+a^2\kappa}.
\end{equation}

\bibliographystyle{apsrev4-1}
\bibliography{RefModifiedGravity.bib}

\end{document}